\documentclass[preprint]{revtex4}
\usepackage{graphicx}
\usepackage{amssymb}
\newcommand{\be}{\begin{eqnarray}}
\newcommand{\ee}{\end{eqnarray}}

\begin{document}

\title{Generalized Parton Distributions for the Proton in Position Space :
Non-Zero Skewness}
\author{\bf R. Manohar$^a$, A. Mukherjee$^a$, D. Chakrabarti$^b$}
\affiliation{$^a$ Department of Physics,
Indian Institute of Technology Bombay,\\ Powai, Mumbai 400076,
India\\
$^b$ Department of Physics, 
Indian Institute of Technology Kanpur,
Kanpur 208016, India}
\date{\today}

\begin{abstract}
We investigate the generalized parton distributions (GPDs) for u and d
quarks in a proton in transverse and longitudinal position space 
using a recent phenomenological parametrization. We take nonzero skewness
$\zeta$ and consider the region $x> \zeta$. Impact parameter space
representation of the GPD $E$ is found to depend sharply on the parameters
used within the model, in particular in the low $x$ region. 
In longitudinal position space a diffraction pattern
is observed, as seen before in several other models.      

\end{abstract}
\maketitle


\vskip0.2in
\noindent
\section{Introduction}

The generalized paton distributions (GPDs) have gained a lot of theoretical
and  experimental interest recently. Unlike the  ordinary parton 
distributions (pdfs) which at a given scale 
depend only on the longitudinal momentum fraction $x$ of the parton, 
GPDs are functions of three variables, $x$, $\zeta$ and $t$ where the 
so-called skewness $\zeta$ gives the
longitudinal momentum transfer and $-t$ is the square of the momentum
transfer in the process. The GPDs give interesting information about the
spin and orbital angular momentum of the quarks and gluons in the nucleon. 
They are experimentally accessed through the overlap of
deeply virtual Compton scattering (DVCS) and Bethe-Heitler (BH) process as
well as exclusive vector meson production \cite{rev}. The real and
imaginary parts of the Compton amplitude give information on GPDs in
different kinematical regions. Several experiments
worldwide, for example, at DESY  HERA collider, by the H1 \cite{H1,H2} 
and ZEUS \cite{ZEUS1,ZEUS2} collaboration and  HERMES \cite{HERMES} fixed 
target experiments have finished taking  data on DVCS. Experiments are also 
being done at JLAB Hall A and B \cite{CLAS}. COMPASS at CERN has programs
to access GPDs through muon beams \cite{COMPASS}. 
A review of the different models and their status with respect to the
experimental results can be found in \cite{boffi}. As the GPDs involve a
momentum transfer (off-forwardness), they do not have probabilistic
interpretation, unlike ordinary parton  distributions (pdfs). However, it
has been shown that when the momentum transfer is purely in the transverse
direction, if one performs a Fourier Transform (FT) with respect to the
transverse momentum transfer $\Delta_\perp$, one gets the so-called impact
parameter  dependent parton distributions (ipdpdfs), originally proposed by
Soper in the context of nucleon form factor \cite{burkardt, soper}. 
Ipdpdfs tell us how the quarks
of a given longitudinal momentum are distributed in the transverse position
or impact parameter space. These obey certain positivity conditions and
unlike the GPDs, have probabilistic interpretation. As transverse boosts are
non-relativistic Galilean boosts in light-front formalism, there is no
relativistic correction to this interpretation, even though we are
considering relativistic systems. Ipdpdfs are  defined in a
proton state with  a sharp plus momentum $p^+$ and localized in the 
transverse plane such that  the transverse center of momentum $R_\perp=0$ 
(normally, one should work with
a wave packet state which is very localized in transverse position space, in 
order to avoid the state to be normalized to a delta function
\cite{diehl}). These give simultaneous information about the longitudinal
momentum fraction $x$ and the transverse distance $b$ of the parton from the
center of the proton and thus give a new insight to the internal structure 
of the proton. A wave packet state  which is transversely polarized is
shifted sideways in impact parameter space. This shift is determined by the
GPD $E$. An interesting interpretation of Ji's sum rule is given in
\cite{bur05} in
terms of ipdpdfs, with $E$ related to the orbital angular momentum carried
by the quarks. Interesting connections between the ipdpdfs and transverse
momentum dependent pdfs have been obtained in various models. However such
relations cannot exist in a model independent way \cite{andreas}. 

Since GPDs depend on  a  sharp $x$, the Heisenberg uncertainty
relation restricts the longitudinal position space
interpretation of GPDs themselves. It has, however, been shown
in \cite{wigner} that one can define a quantum mechanical Wigner
distribution for the relativistic quarks and gluons inside the proton.
Integrating over $k^-$ and $k^\perp$, one
obtains a four dimensional quantum distribution which is a function
of ${\vec{r}}$ and $k^+$ where  ${\vec{r}}$ is the quark phase space
position defined in the rest frame of the proton. These
distributions are related to the FT of GPDs in the same frame. This gives a
3D position space picture of the GPDs and of the proton, within the
limitations mentioned above. In \cite{metz}, a parametrizations of generalized
parton correlation functions have been done for a spin $1/2$ target. These
reduce to GPDs when integrated over $k_\perp$ and to TMDs when the momentum
transfer is zero.

In another work, the real and imaginary parts of the DVCS amplitude are
expressed in longitudinal position space by introducing a
longitudinal impact parameter $\sigma$ conjugate to the skewness $\zeta$.
Taking a field theory inspired simple relativistic spin $1/2$ system, namely
for an electron dressed with a photon in QED, it was found that the DVCS
amplitude show a diffraction-like pattern in longitudinal position space
\cite{hadron_optics}.
Since Lorentz boosts are kinematical in the front form, the correlation
defined in the 3 D position space $b_\perp$ and $\sigma$ is frame
independent. Similar diffraction pattern was observed in a holographic model
for the meson. In this work, we investigate a recent parametrization
\cite{liuti2} of
the GPDs in position space for non-zero $\zeta$. GPDs for zero skewness have
been parametrized in  a similar way in \cite{liuti}. At the input scale 
they are
parametrized by a spectator model term multiplied by a Regge motivated term. 
The parameters were fitted by fitting the forward pdfs and form factors.
For non-zero $\zeta$, the GPDs have to satisfy an additional constraint, 
namely polynomiality. In certain models, for example using the overlap of
light-front wave functions (LFWFs), it is very difficult to obtain a
suitable parametrization of the higher Fock components of the wave function 
in order to get the polynomiality of GPDs. Polynomiality is
satisfied by construction only if one considers the LFWFs of simple spin 
$1/2$ objects like a dressed quark or a dressed electron  in perturbation
theory instead of the proton \cite{overlap,us}. A recent fit to the DVCS
data at small Bjorken $x$ from H1 and ZEUS was done in \cite{dieter}, 
using the conformal Mellin-Barnes representation of the DVCS amplitude. 
However, to get the GPDs one has to do an inverse Mellin transform, and a 
knowledge of all moments are required for that. In \cite{liuti2}  a functional
form of GPDs in the  DGLAP region $x> \zeta$ has been obtained by 
generalizing a parametrization for zero skewness. In the ERBL region 
$x< \zeta$, a quark and an antiquark pair emerges from the nucleon and
undergoes the electromagnetic interaction. Lattice moments were used as
constraints and weighted average of GPDs around definite values of $x$ were
constructed using Bernstein polynomials. However, so far only a few of the
lattice moments are known and this method has large theoretical
uncertainties.  

In a previous publication \cite{us3} we used the GPD parametrization  of 
\cite{liuti} at zero skewness to calculate the distributions of partons in
the transverse impact parameter space. In this work, we study the GPDs in
transverse and longitudinal position space using the phenomenological 
parametrization  in \cite{liuti2} for nonzero skewness.  
The plan of the paper is as follows. After presenting the model used in
section II, we discuss numerical results for the GPDs in transverse and
longitudinal position space respectively in section III. Summary and
conclusions are given in section IV.    

\vskip0.2in
\noindent
\section{Parametrization of the GPDs} 

We consider the parametrization in \cite{liuti,liuti2} for the GPDs for 
nonzero skewness ($\zeta\ne 0$) :

\noindent {\bf Set I} 
\begin{eqnarray}
H^I(x,\zeta,t) & = & G_{M_{x}^I}^{\lambda^I}(x,\zeta,t) \,  
 x^{-\alpha^I -\beta_1^I (1-x)^{p_1^I} (t+t_{min})}
\label{param1_H}
\\
E^I(x,\zeta,t) & = & \kappa \, G_{M_x^I}^{\lambda^I}(x,\zeta,t) \,
x^{-\alpha^I -\beta_2^I (1-x)^{p_2^I}( t+t_{min})}
\label{param1_E}
\end{eqnarray}

\noindent {\bf Set II} 
\begin{eqnarray}
H^{II}(x,\zeta,t)& = & G_{M_x^{II}}^{\lambda^{II}}(x,\zeta,t) \,
x^{-\alpha^{II} - \beta_1^{II}  (1-x)^{p_1^{II}} (t+t_{min})}
\label{param2_H}
\\
E^{II}(x,\zeta,t) & = &
G_{\widetilde{M}_x^{II}}^{\widetilde{\lambda}^{II}}(x,\zeta,t) \,
x^{-\widetilde{\alpha}^{II} - \beta_2^{II}  (1-x)^{p_2^{II}} (t+t_{min})}
\label{param2_E}
\end{eqnarray}

All parameters except for $p_1$ and $p_2$ are flavor dependent.
The function $G$ has the same form for both parametrizations, I and II:

\begin{eqnarray}
G_{M_x}^{\lambda} (x,\zeta,t) = && 
{\cal N} \frac{x}{1-x} \int d^2{\bf k}_\perp \frac{\phi(k^2,\lambda)}{D(x,{\bf k}_\perp)}
\frac{\phi({k^{\prime \, 2},\lambda)}}{D(x,{ \bf k}_\perp -\frac{1-x}{1-\zeta}{\bf 
\Delta}_\perp)},    \label{gkaava}
\end{eqnarray}
where

\begin{equation}
D(x,{\bf k}_\perp) \equiv  k^2 - m^2,
\end{equation}

\begin{eqnarray}
k^2 & = & x M^2 - \frac{x}{1-x} M_x^{2}  - \frac{{\bf k}_\perp^2}{1-x} \\
k^{\prime \, 2} & = & xM^2 - \frac{x}{1-x} M_x^{2} - \Big({\bf k}_\perp 
- \frac{1-x}{1-\zeta}\Delta\Big)^2~\frac{1-\zeta}{1-x},
\end{eqnarray}
and 
\begin{eqnarray}
\phi(k^2,\lambda) = \frac{k^2-m^2}{\vert k^2-\lambda^2\vert^2 },
\label{phi}
\end{eqnarray}
When the skewness is nonzero, the total momentum transfer square is modified to :
\be
-t=\Delta^2=\frac{4\xi^2M^2}{1-\xi^2}+(1-\xi^2)D^2=-t_{min}+(1-\xi^2)D^2
\ee
where $D=P^\prime/(1-\xi)-P/(1+\xi)$ and $\xi=\zeta/(2-\zeta)$.  
$D$ reduces to $\Delta$ at $\zeta=0$.

Here $x$ is the fraction of the light
cone momentum carried by the active quark, $k$ being its momentum. 
The mass parameters are  $m$, the struck quark mass, and $M$, the proton mass.
The normalization factor includes the nucleon-quark-diquark coupling, and it
is  set to ${\cal N} = 1$ GeV$^6$.  Here we consider the DGLAP region 
$\zeta<x<1$. The dominating contribution in this  kinematical region 
comes from the process where a quark from the proton with momentum 
fraction $x$ is struck by the incident  photon and again reabsorbed by 
the proton. The above phenomenologically motivated parametrization of the
GPDs $H(x,\zeta, t)$ and $E(x,\zeta, t)$ at zero skewness 
$\zeta$ was done in \cite{liuti} using a
spectator model calculation at the  low input scale. The
spectator model has been used for its simplicity and for the fact that it  
is flexible enough to predict the main features of a number of distribution
and fragmentation functions in the intermediate and large $x$ region. The
spectator mass is chosen to be different for different quark flavor GPDs.
However, similar to the case of pdfs, the spectator model is not able to
reproduce quantitatively the small $x$ behaviour of the GPDs. So a   
`Regge-type' term has been considered multiplying the spectator model
function $G_{M_x}^\lambda$. Extension to nonzero $\zeta$ was done in
\cite{liuti2}. 
 The parameters are listed in \cite{liuti} for both the sets.
The parameters  $M_{x}^q$, $\lambda^q$ and $\alpha^q$, $q=u,d$, obtained
at an initial scale $Q_0^2$ ($Q^2_0 = 0.094$ GeV$^2$),
and they are the same for both Sets I and II, in Set I they are by 
definition the same for the functions $H$ and  $E$ 
(see Eqs.~(\ref{param1_H},\ref{param1_E})). 
The parameters $\beta_1$, $\beta_2$, ${p_1}$ and ${p_2}$, in Set I, and 
all parameters defining $E$ in Set II (Eq.~({\ref{param2_E}}),
were fitted  to the nucleon electric and magnetic form factors, 
with the values of $M_x^q$, $\lambda^q$, and $\alpha^q$ fixed. The input
scale $Q_0^2=0.094 {\mathrm{GeV}}^2$ is obtained as a parameter in the
model. The low value of $Q_0^2$ results from the requirement that only
valence quarks contribute in the momentum sum rule. The GPD H is constrained
in the forward limit by the pdf data. As the GPD $E$ is
unconstrained by the data on forward pdfs, two different sets of parameters
were used in the fit, these are denoted by set I and II. In set II an additional
normalization condition was used
\be
\int_0^1 dx E_q(x,t=0)=\kappa^q
\ee
where $\kappa^u$ and $\kappa^d$ are the u and d quark contributions to the
nucleon anomalous magnetic moment. Although $H^u$ and $H^d$ have similar
behaviour in both the sets, $E^u$ and $E^d$ behave differently even for
$\zeta=0$. In the  forward limit for H,  Alekhin \cite{ale} leading order pdf 
sets were   used. Additional constraints for nonzero $\zeta$ were
obtained from lattice moments.

\vskip0.2in
\noindent
\section{Parton distributions in impact parameter space}

Parton distribution in impact parameter space is defined
as :
\be
q(x,\zeta,b)={1\over 4 \pi^2} \int d^2 \Delta_\perp e^{-i \Delta_\perp 
\cdot b_\perp} H(x,\zeta,t)\nonumber\\
e(x,\zeta,b)={1\over 4 \pi^2} \int d^2 \Delta_\perp  e^{-i \Delta_\perp 
\cdot b_\perp} E(x,\zeta, t).
\ee


Here $b= \mid b_\perp \mid $ is the transverse  impact parameter which is a 
measure of  the transverse distance between
the struck parton and the center of momentum of the hadron.  $b$ is defined 
such a way that $\sum_i x_i b_i=0$ where the sum is over the number of partons.
The relative distance ${b\over 1-x}$ between the struck parton and the
spectator system provides an estimate of the size of the system as a whole.
The above picture was proposed in the limit of zero skewness $\zeta$ in
\cite{burkardt}. In most experiments $\zeta$ is nonzero, and it is
of interest to investigate the GPDs in $b_\perp$ space with  
nonzero $\zeta$. 

As described in the introduction the ipdpdfs describe the probability of
finding  a parton of definite momentum fraction $x$ at a distance $b_\perp$
from the center of the proton. When the skewness $\zeta$ is nonzero, the
transverse location of the proton itself is different before and after
the scattering. This transverse shift does not depend on $x$ but on the
skewness $\zeta$ and $b= \mid b_\perp \mid$. The information on the
transverse shift is not washed out even if the GPDs are integrated over
$x$ in the DVCS amplitude. In the DGLAP region
$x>\zeta$, the impact parameter $b_\perp$ gives the location where the
quark is pulled out and put back to the nucleon. In the ERBL region $\zeta <
x$, $b_\perp$ denotes the transverse location of the quark-antiquark pair
inside the nucleon. For a single fermion, the impact parameter dependent pdf
would be a delta function. The smearing in $b_\perp$ space is due to the
multiparticle correlation. In Fig. 1 we plot $e_u(x, \zeta, b)$ and 
$e_d(x, \zeta, b)$ as a function of $b$ for a fixed value of $x$ and
different values of $\zeta$. Substantial difference is seen in the impact
parameter space for the two sets of parametrization, set I and set II. In
set I both $e_u$ and $e_d$  become more sharply peaked at $b=0$ when $\zeta
$ is smaller. It is to be noted that 
the change in $e(x,b)$ with $b$ is related to the deformation of the parton
distribution in impact parameter space for transversely polarized nucleon.
For non-zero $\zeta$, this probes the deformation when the nucleon has a
transverse shift before and after scattering.
 In set II for $e_u$, the peak increases as
$\zeta$ increases, in contrast to set I, whereas for $e_d$, the peak
decreases for increasing $\zeta$ but does not become broader as in set I.

In Fig. 2, we have plotted $q(x,\zeta,b)$ as a function of $b$ for fixed
$x=0.6$ and different values of $\zeta$. As the two parametrizations set I
and II are not much different for the GPD $H(x,\zeta,t)$ not much difference
is seen in the impact parameter space. In all cases here the peak in $q(x,
\zeta,b)$ for fixed $x$ decreases with increase of $\zeta$. For d quark
distributions, the peak also becomes broader as $\zeta$ increases, which
is the same behaviour as seen in $e(x,\zeta,b)$. As $ \zeta$ increases 
for fixed $x$, the smearing in $b$ space
becomes broader, which means that as the momentum transfer in the
longitudinal direction increases, the active quark is more likely to be
pulled out at a larger $b$. 

In Figs. 3 and 4 we plotted the impact parameter dependent distributions
$e(x, \zeta, b)$ and $q(x, \zeta,b)$ for fixed $\zeta$ and $b$ values as
functions of $x$. Substantial difference is observed in $e(x, \zeta, b)$  
between set I and II, in particular in the low $x$ region. For $e_u$ the
peak shifts towards smaller $x$ values ($x> \zeta$ in our calculation).  
However for $-e_d$ the peak shifts to higher $x$ values in set II. The
qualitative behaviour of $q_u$ and $q_d$ are the same as functions of $x$.

The boost invariant  longitudinal impact parameter $\sigma$ was first introduced 
in \cite{hadron_optics} and it was shown that DVCS amplitude shows interesting diffraction 
pattern in longitudinal impact parameter space.
GPDs also when expressed in term of $\sigma$ exhibit the similar diffraction 
pattern \cite{us2}. The boost invariant longitudinal impact parameter conjugate 
to the longitudinal momentum transfer is defined as $\sigma=\frac{1}{2}b^-P^+$.  
So, the GPDs in longitudinal position space is given by:
\be
q(x,\sigma, t)&=&\frac{1}{2\pi}\int_0^{\zeta_f} d\zeta e^{i \zeta P^+b^-/2} H(x,\zeta,t)\nonumber\\
&=& \frac{1}{2\pi}\int_0^{\zeta_f} d\zeta e^{i \zeta \sigma} H(x,\zeta,t)\nonumber\\
e(x,\sigma, t)&=&\frac{1}{2\pi}\int_0^{\zeta_f} d\zeta e^{i\zeta P^+b^-/2} E(x,\zeta,t)\nonumber\\
&=& \frac{1}{2\pi}\int_0^{\zeta_f} d\zeta e^{i \zeta \sigma} E(x,\zeta,t)
\ee
Since we are concentrating only in the region $\zeta<x<1$, the upper 
limit of $\zeta$ integration $\zeta_f$ is given by $\zeta_{max}$ 
if $x$ is larger then $\zeta_{max}$, otherwise by $x$ if $x$ is smaller than
$\zeta_{max}$ where $\zeta_{max}$ is the maximum value of $\zeta$ allowed 
for a fixed $-t$: 
\be
\zeta_{max}={(-t)\over 2 M^2}\Big( \sqrt{1+{4 M^2\over (-t)}}-1 \Big).
\ee 

In Figs. 5 and 6 we have
plotted the GPDs in longitudinal position space $\sigma$. 
We restrict ourselves to the DGLAP region. The GPDs show diffraction 
pattern in $\sigma$
space, similar to that observed for a dressed electron in QED or in a
holographic model for the meson \cite{hadron_optics}. There is a primary 
maximum followed by a
series of secondary maxima. The positions of the minima are the same for
both u and d quark GPDs and they do not depend whether we are plotting $                  
q(x,\sigma,b)$ or $e(x,\sigma,b)$. The feature was also observed for a
dressed electron state in \cite{hadron_optics}. The positions of the 
minima are characteristics  of
the finite Fourier transform and independent of the GPD used. As $-t$
increases, the positions of the first minima move in to smaller values of
$\sigma$. As seen in Fig. 5, for set II, the peaks in $e_u(x,\sigma,t)$   
are more sharp compared to set I. On the other hand, for
$e_d(x,\sigma,t)$ the magnitude of the peak is more affected as $-t$
increases in set II compared to set I. However for $q_u(x,\sigma,t)$ and
$q_d(x,\sigma,t)$ there is not much difference between the two
parametrization sets, as expected. In all cases, the magnitude of the peak
decreases as $-t$ increases and the first minima move in. This effect has
already been observed for a dressed electron state and in  a holographic
model for the meson \cite{hadron_optics}; as well as for chiral odd GPDs
\cite{us2}. In \cite{hadron_optics} a relation
between the position of the first minima and the momentum transfer squared
$-t$ has been derived in analogy with diffraction in optics. The fact that
such pattern with the same qualitative behaviour is observed here in a
phenomenological parametrization of GPDs shows that it is not an artefact of
the cuts and constraints used in the field theory based model. In
\cite{pion} pion distribution amplitudes (DAs) have been investigated in
longitudinal position space (Ioffe time) by expressing the DAs in an
expansion over Gegenbauer polynomials.  Oscillation of the DA has been
observed in longitudinal position space showing two humps in certain models.
It was concluded that this is an artefact of truncating the conformal spin
partial wave expansion. In order to further investigate the diffraction
pattern that we observe for the GPDs, in Fig. {\ref{model12}} we plot the GPD $H$
in longitudinal position space using a model similar to that in \cite{marc}.
For the dotted and long dashed curves we chose a factorized ansatz of the form
\be 
H(x, \zeta, t)= N u(x) F_1(t)
\label{model1}
\ee
where $N$ is the normalization constant, we took $u(x)=x (1-x) $ and for the
form factor $F_1(t) $ we took the standard dipole form. The FT shows
diffraction pattern with the general features the same as observed in other
models. In the solid and short-dashed curves, we used a somewhat different ansatz,
namely
\be
H(x, \zeta, t)= {\cal{N}} u(x,\zeta) F_1(t)
\label{model2}
\ee
where we took the same dipole form factor, and $u(x, \zeta)= (x-\zeta)^2
(1-x)^2$; ${\cal{N}}$ being the normalization constant. The normalization
constants have been chosen to fit the two sets of curves in the same plot.
There is no diffraction pattern in the second model; which clearly shows
that the pattern is not only due to the finite range of the $\zeta $ integration
but depends also on the $x, \zeta$ and $t$ interplay in  the GPD
parametrization used. So an experimental observation of such diffraction
pattern can help to constrain GPD parametrizations. However in order      
to get the full Lorentz invariant picture in longitudinal position space one
has to consider the other kinematical region $x<\zeta$ as well; as explained
in \cite{hadron_optics}. Also the positions of the minima and their shift
with increase of $-t$ are model independent features of the Fourier
transform.  

\vskip0.2in
\noindent
{\section{Conclusion}}
In this work we investigated the GPD for u and d quark distributions in the
proton using a recent parametrization \cite{liuti2}. For nonzero $\zeta$ we
worked in the DGLAP region $x> \zeta$. Taking a Fourier transform with
respect to the transverse momentum transfer $\Delta_\perp$ we obtained the
parton distributions in impact parameter space. For nonzero $\zeta$ these
probe the parton distributions when the initial proton is shifted from the
final proton in the transverse plane. When the proton is transversely
polarized, the parton distributions in the tranverse plane is distorted,
this distortion is related to the GPD $E(x,\zeta,t)$ and to the orbital
angular momentum of the quarks. We showed the $x$ and $b_\perp$ dependence
of the ipdpdfs for various $\zeta$ values. We introduced a boost invariant
longitudinal impact parameter $\sigma$ conjugate to $\zeta$. Both the GPDs            
$H$ and $E$ in $\sigma$ space show diffraction pattern as seen before in some
other models. We did some comparative study using different toy models and
showed that although the general features of this pattern are independent of
specific models but the appearance of the diffraction pattern depends not
only on the finiteness of the $\zeta$ integration but also on the interplay
of the $x$, $\zeta$ and $t$ dependence of the GPDs.  
\vskip0.2in
\noindent
\section{Ackowledgement:} 
AM thanks DST fasttrack scheme, Govt. of India, for support. 



 \newpage


\begin{figure}[htbp]
\begin{minipage}[c]{0.9\textwidth}

\tiny{(a)}\includegraphics[width=6.5cm,height=5cm,clip]{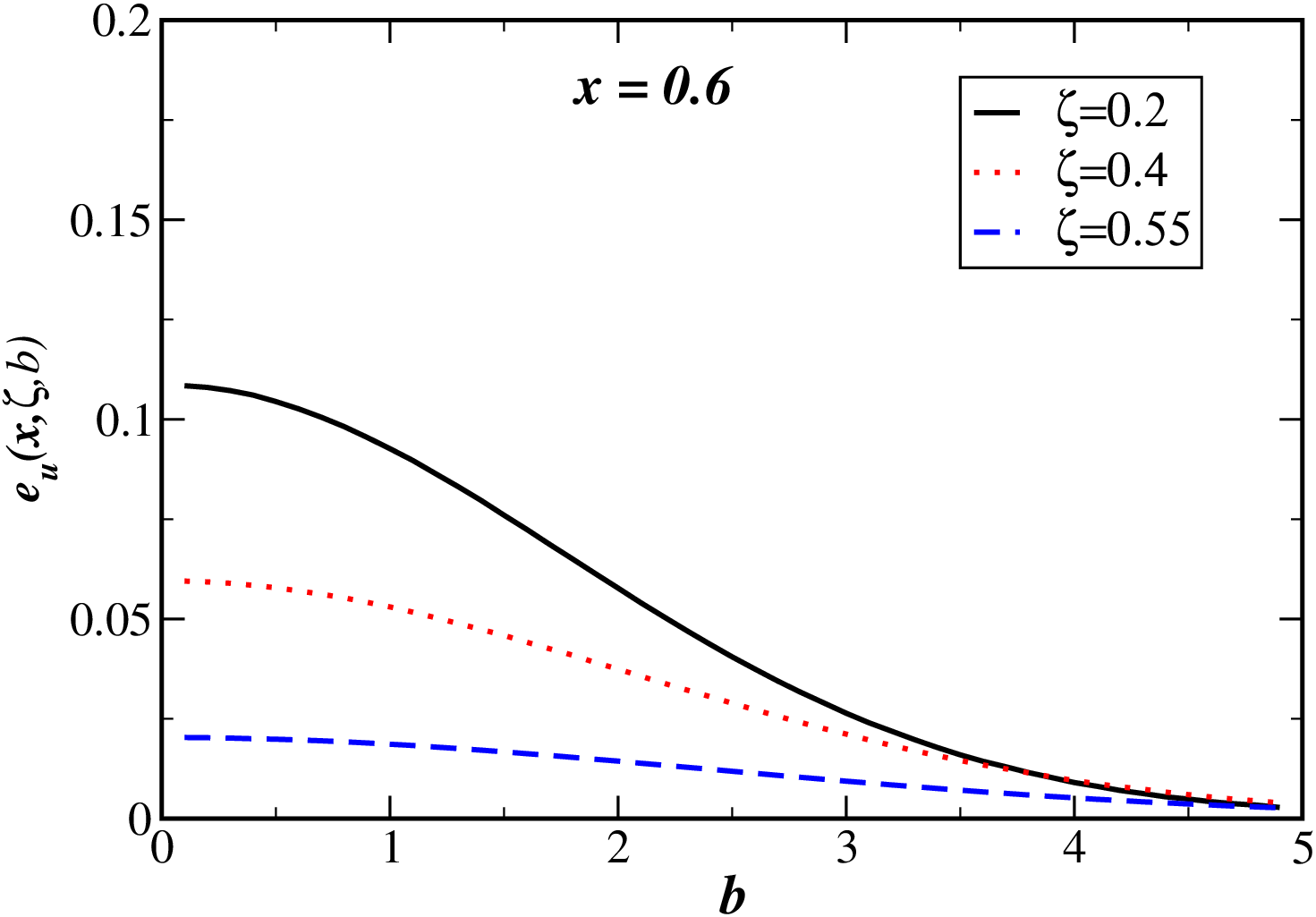}
\hspace{0.1cm}
\tiny{(b)}\includegraphics[width=6.5cm,height=5cm,clip]{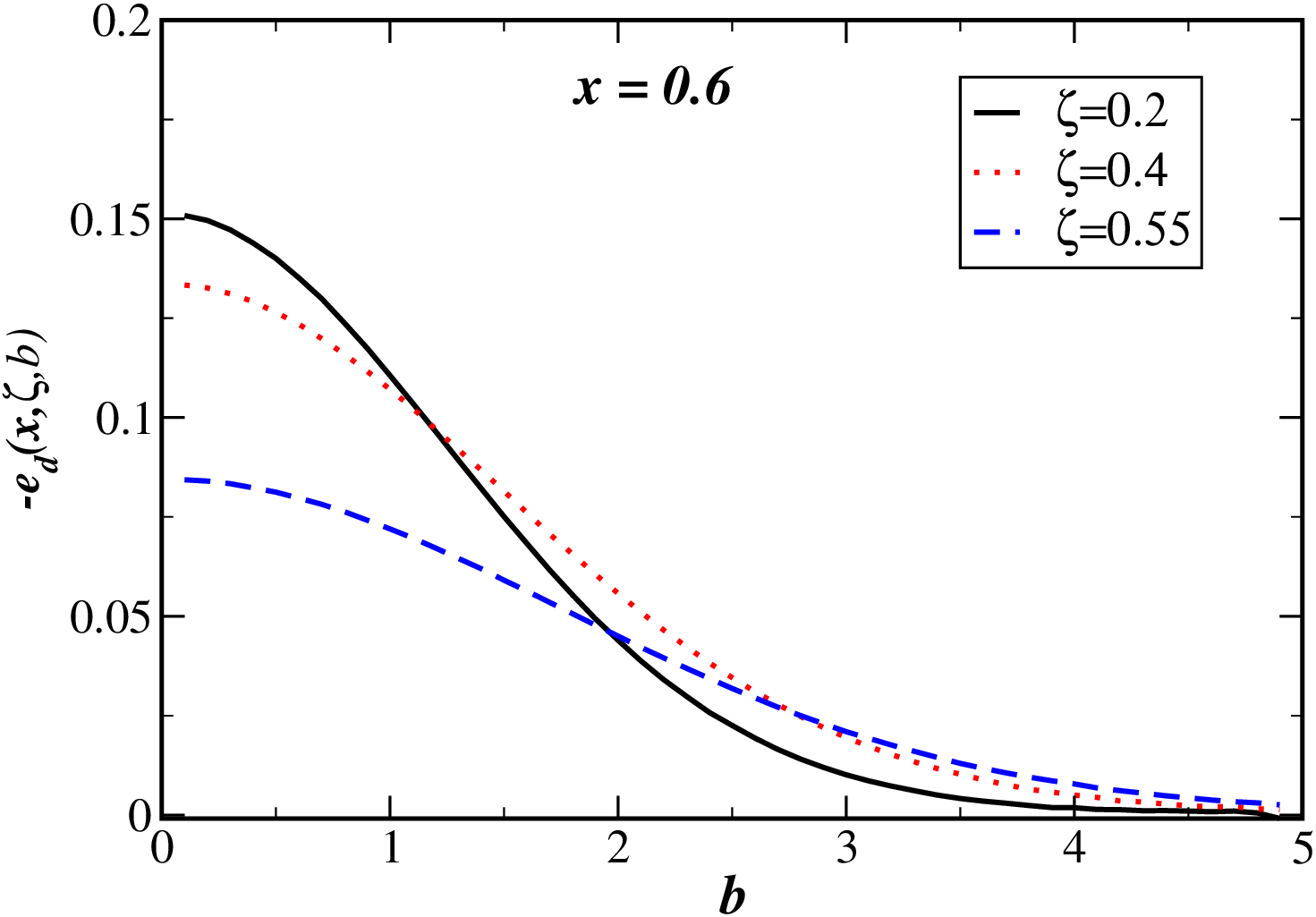}
\end{minipage}
\begin{minipage}[c]{0.9\textwidth} 
\tiny{(c)}\includegraphics[width=6.5cm,height=5cm,clip]{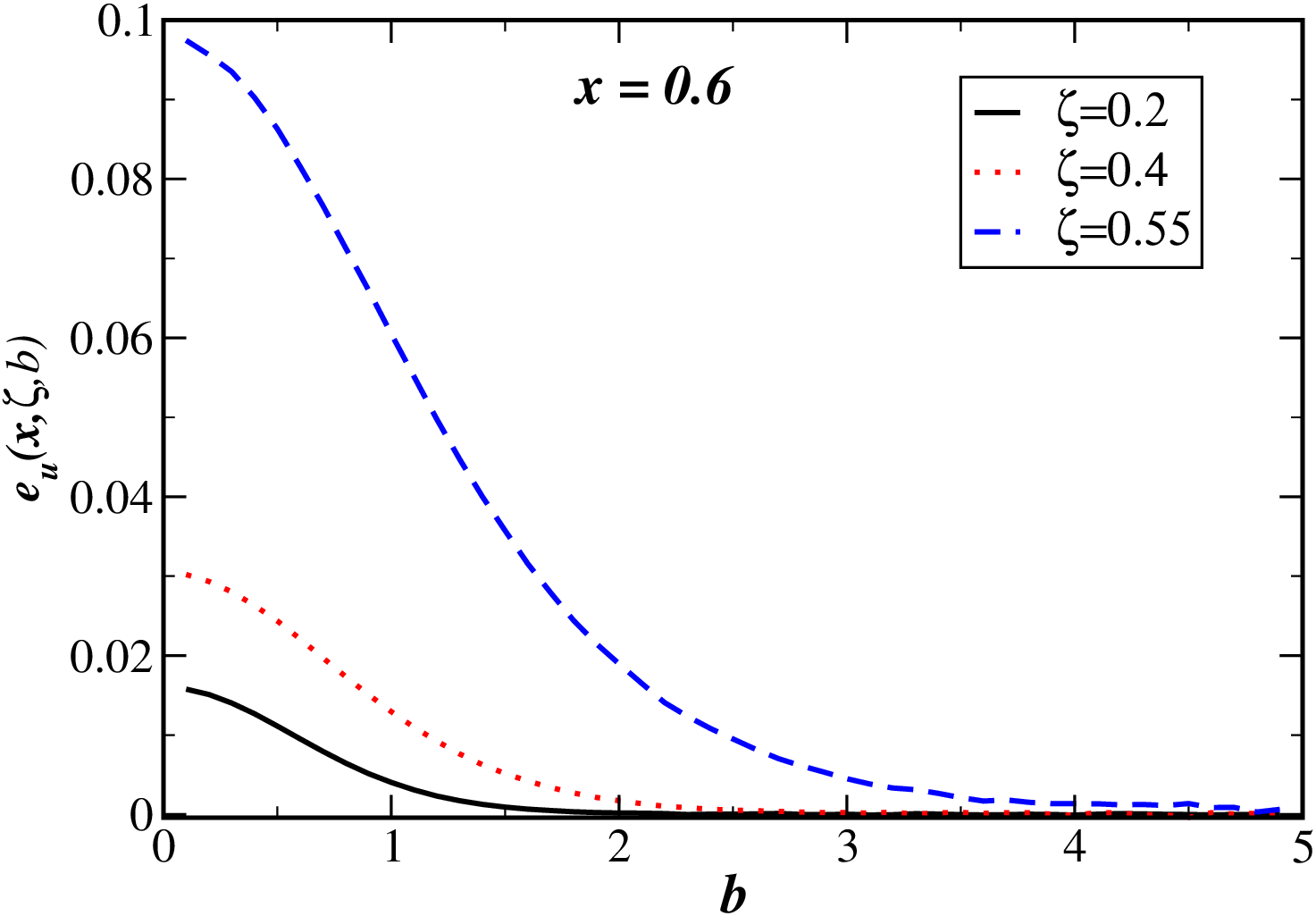}
\hspace{0.1cm}%
\tiny{(d)}\includegraphics[width=6.5cm,height=5cm,clip]{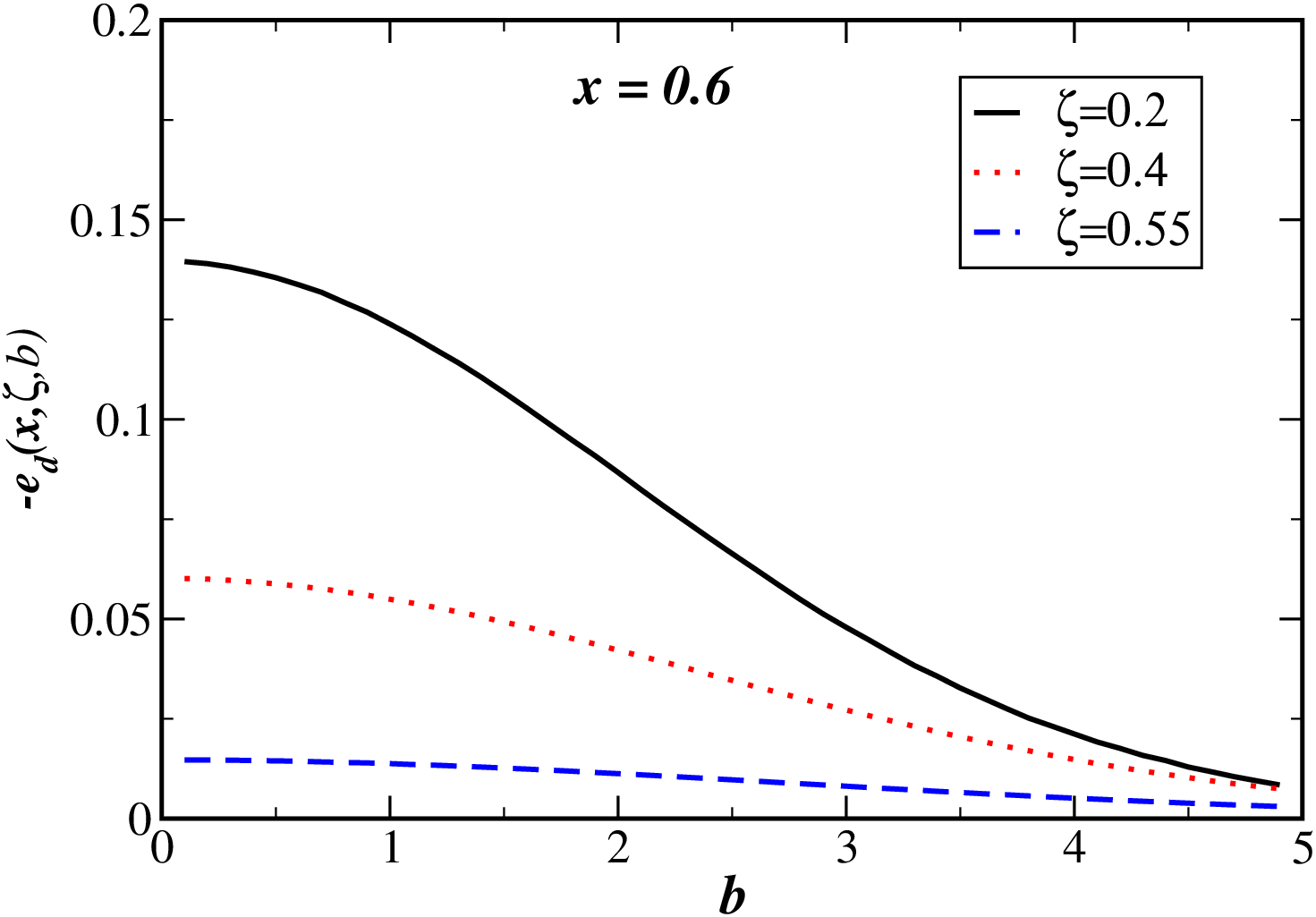}
\end{minipage}
\caption{\label{ipd-eud}(Color online) Plots of (a) $e_u(x,\zeta,b)$ 
vs $b = \mid b_\perp \mid$ for fixed values of $x$ and $\zeta$ and for parameters as in set I 
 (b) same as in (a) but for $d$ quark, (c) same as in (a) but the parameters are as in 
set II, (d) same as in (b) but the parameters are as in set II. $b$ is in
GeV$^{-1}$.}
\end{figure}

\begin{figure}[htp]
\begin{minipage}[c]{0.9\textwidth}
\tiny{(a)}\includegraphics[width=6.5cm,height=5cm,clip]{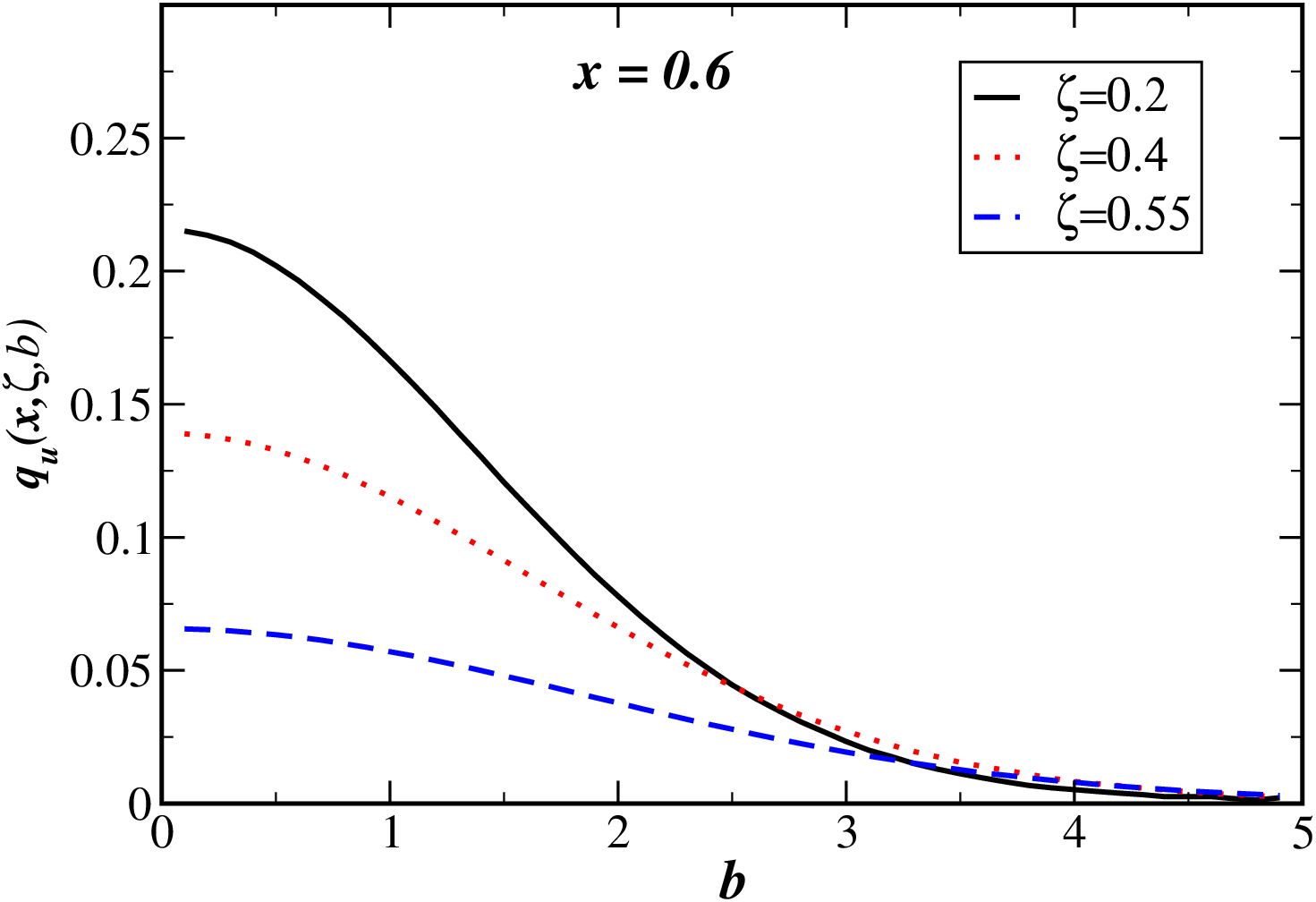}
\hspace{0.1cm}
\tiny{(b)}\includegraphics[width=6.5cm,height=5cm,clip]{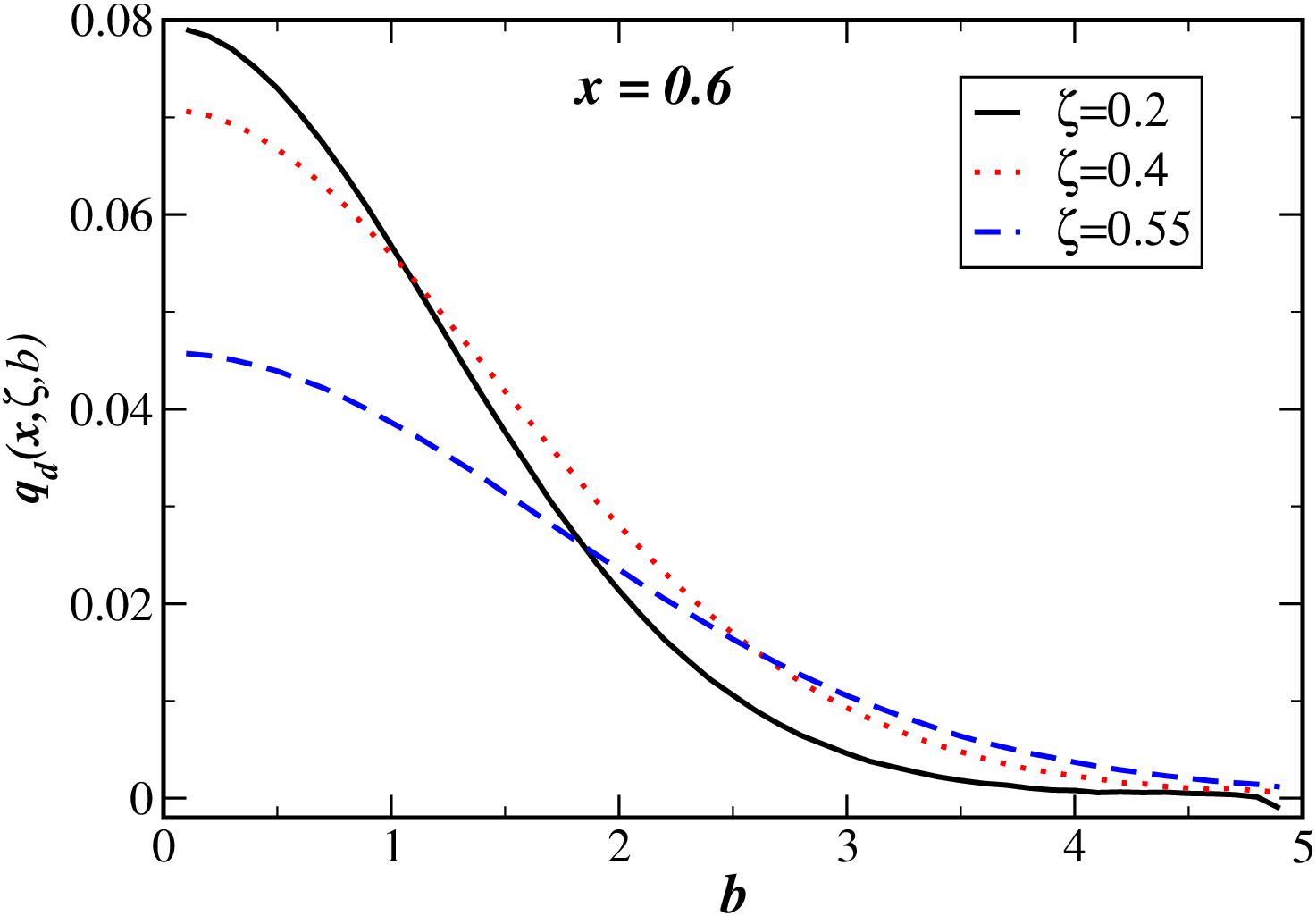}
\end{minipage}
\begin{minipage}[c]{0.9\textwidth} 
\tiny{(c)}\includegraphics[width=6.5cm,height=5cm,clip]{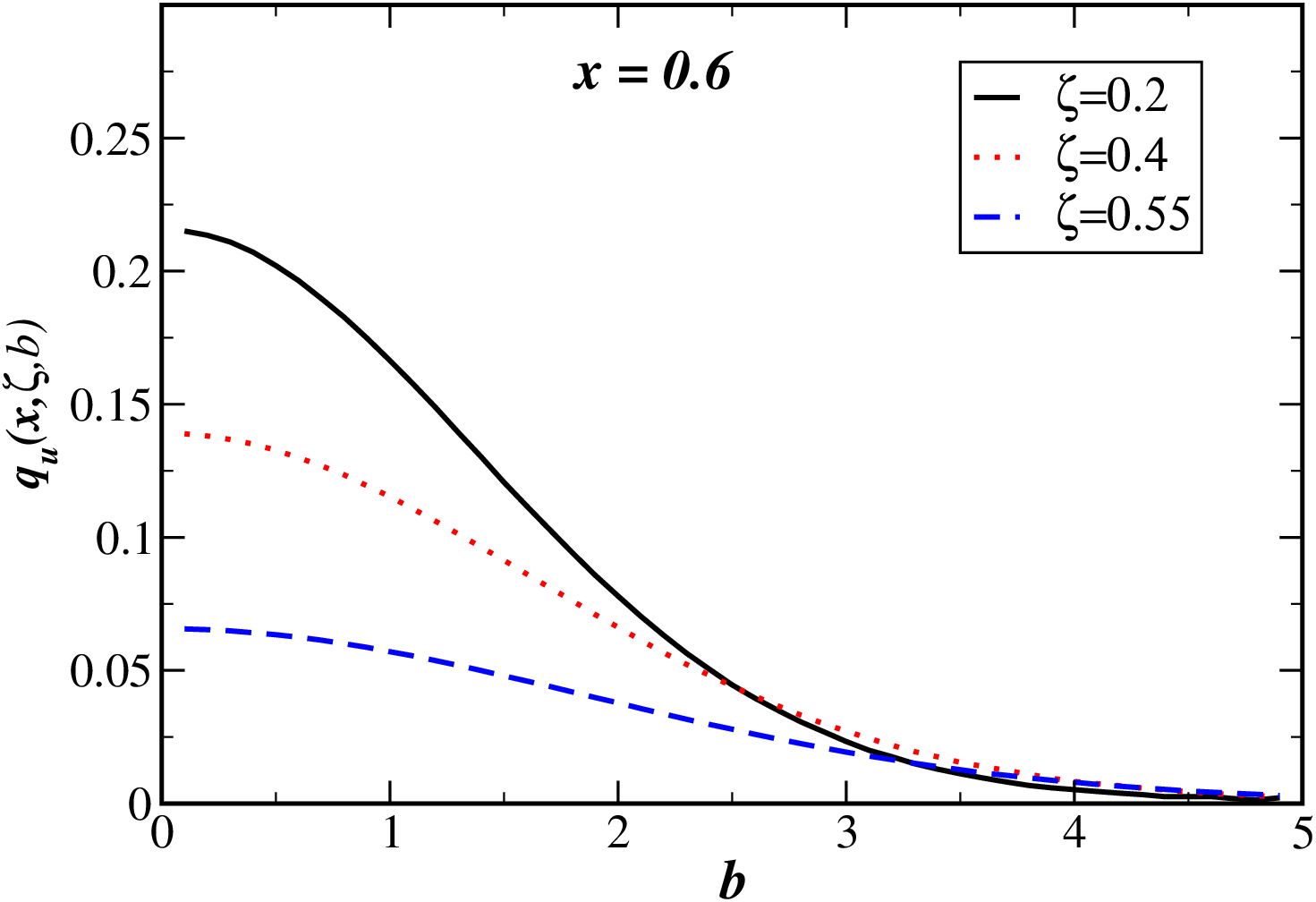}
\hspace{0.1cm}%
\tiny{(d)}\includegraphics[width=6.5cm,height=5cm,clip]{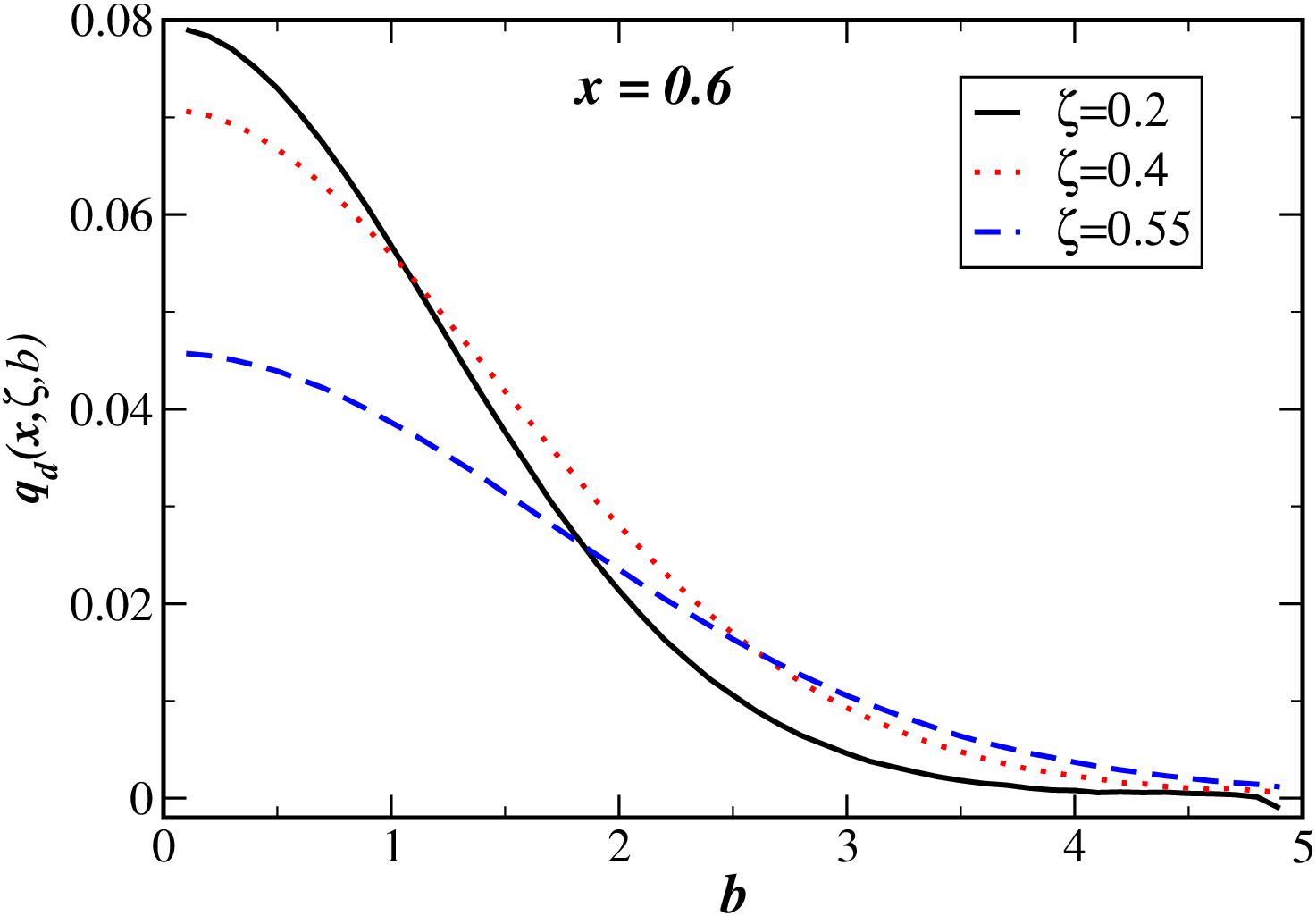}
\end{minipage}
\caption{\label{ipd-hud}(Color online) Plots of (a) $q_u(x,\zeta,b)$ vs $b  = \mid b_\perp \mid$ for
fixed values of $x$ and $\zeta$ and for parameters as in set I,  (b) same as in (a) but for $d$ quark, 
(c) same as in (a) but the parameters are as in set II, (d) same as in (b) but for
parameters as in set II. $b$ is in GeV$^{-1}$. }

\end{figure}

\begin{figure}[htp]
\begin{minipage}[c]{0.9\textwidth}
\tiny{(a)}\includegraphics[width=6.5cm,height=5cm,clip]{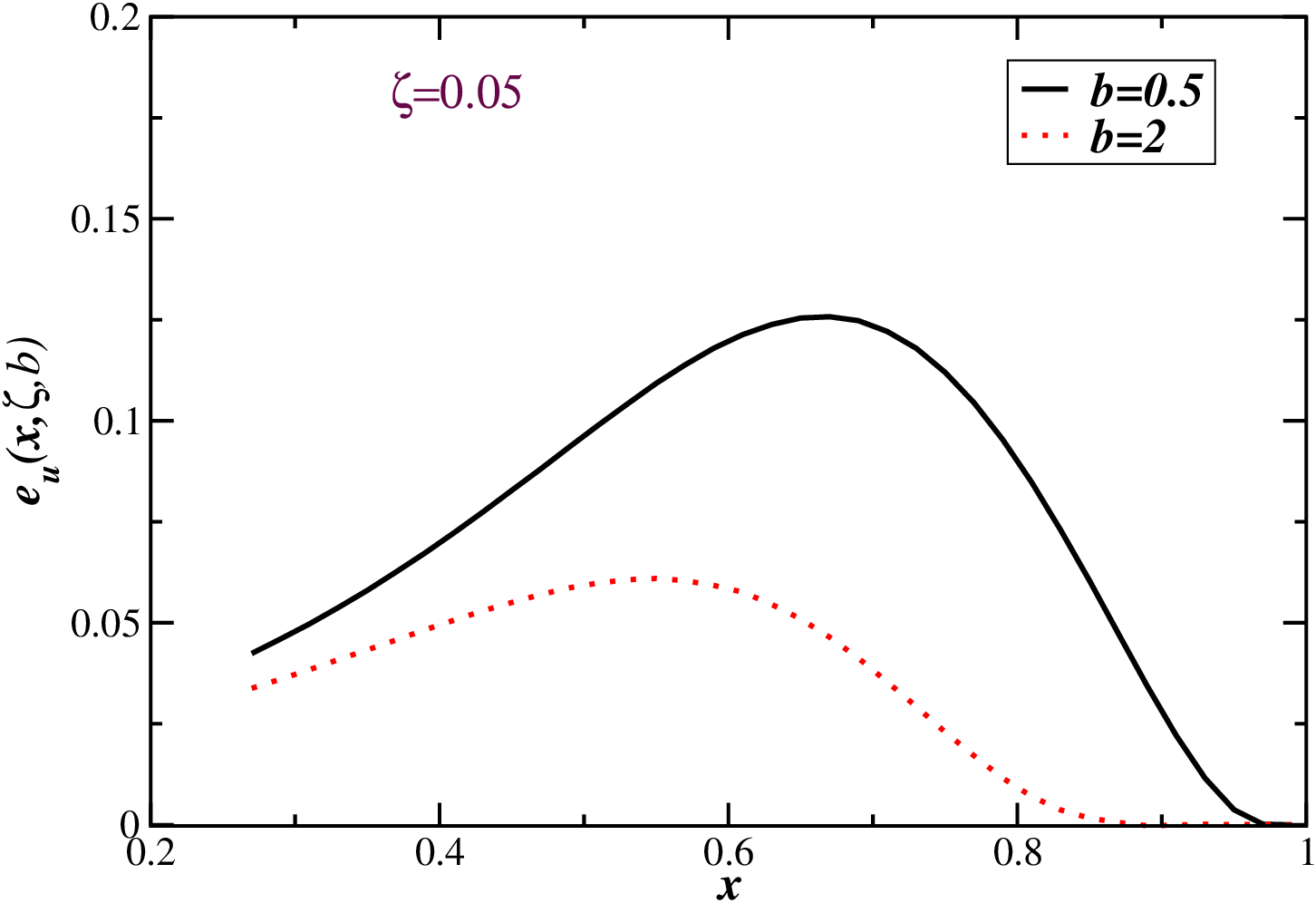}
\hspace{0.1cm}
\tiny{(b)}\includegraphics[width=6.5cm,height=5cm,clip]{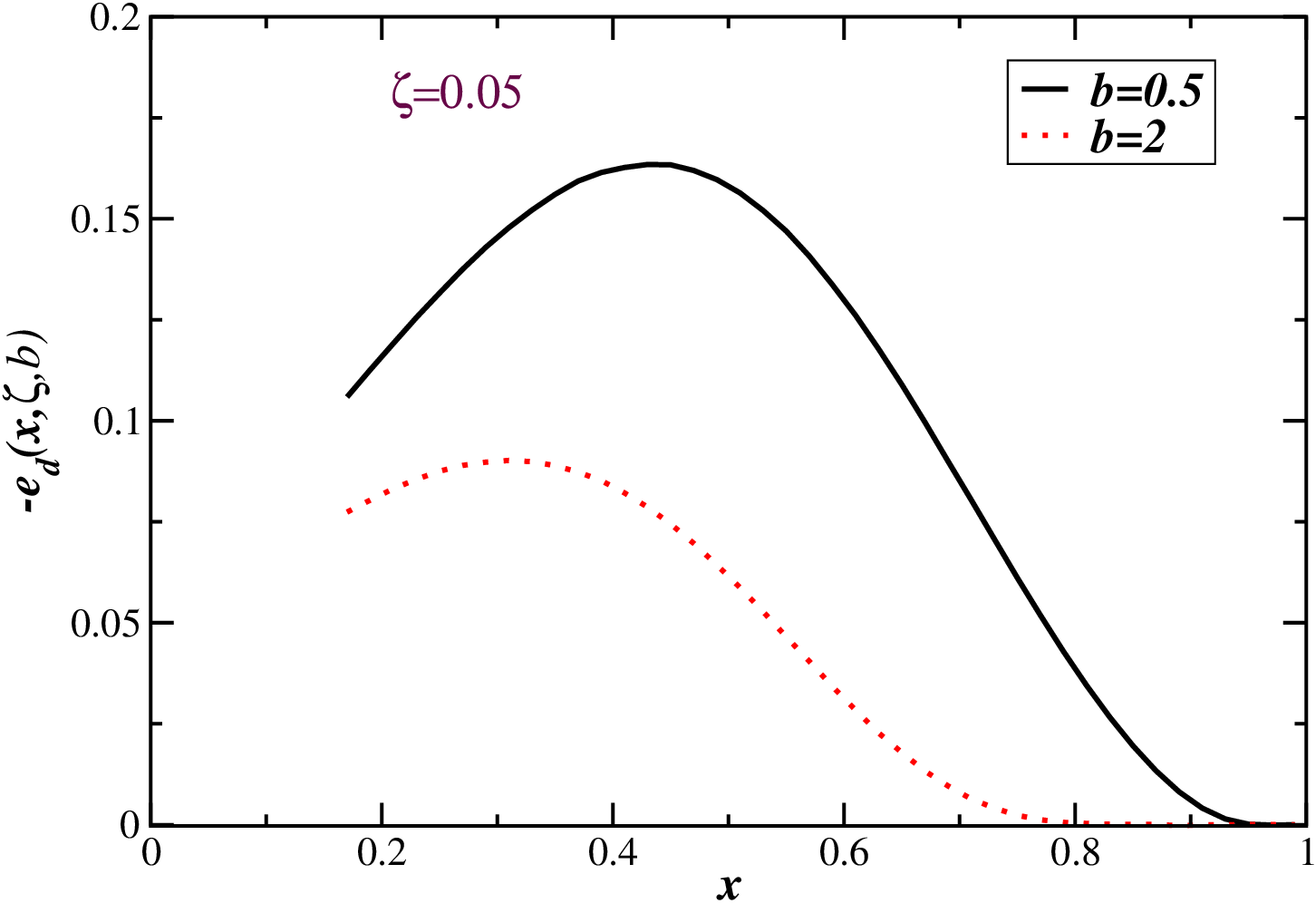}
\end{minipage}
\begin{minipage}[c]{0.9\textwidth}
\tiny{(c)}\includegraphics[width=6.5cm,height=5cm,clip]{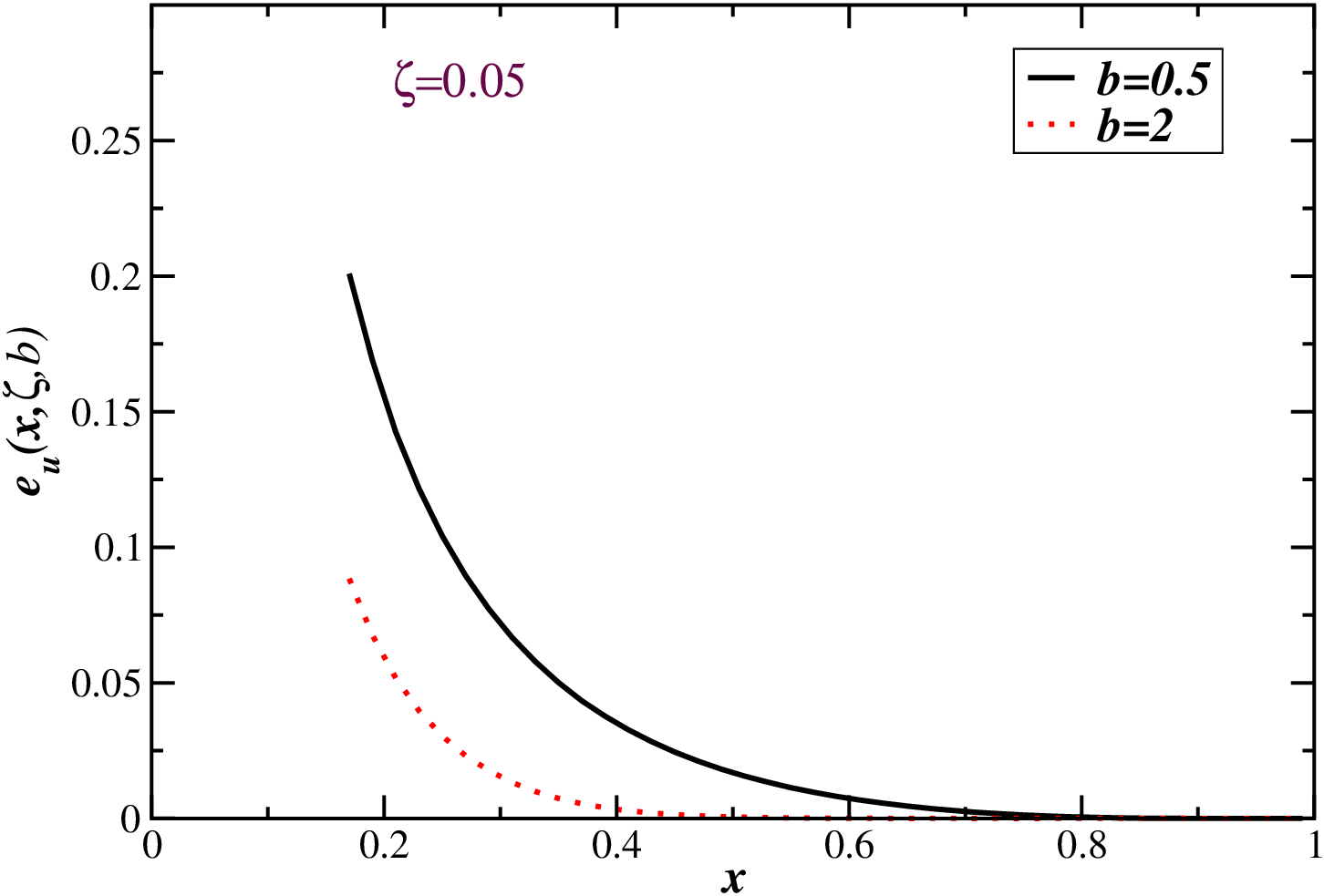}
\hspace{0.1cm}%
\tiny{(d)}\includegraphics[width=6.5cm,height=5cm,clip]{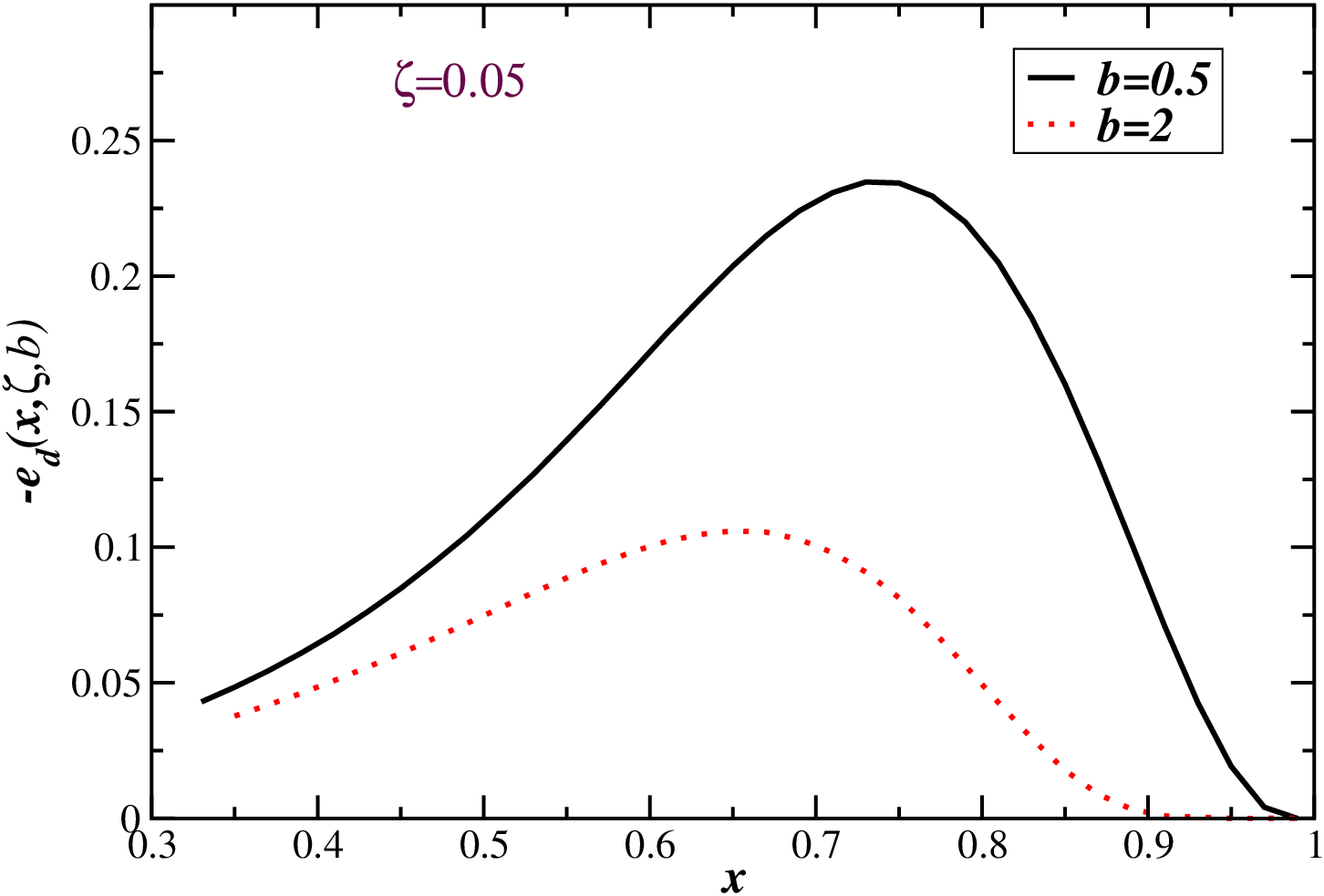}
\end{minipage}
\caption{\label{ipde-x}(Color online) Plots of (a) $e_u (x,\zeta,b)$ 
vs $x$
for fixed values of $b$ and $\zeta$ and for parameters as in set I,
(b) same as in (a) but for $d$ quark, (c) same as
in (a) but for parameters as in set II, (d) same as in (b) but for
parameters as in set II.  $b$ is in
GeV$^{-1}$.}

\end{figure}
\begin{figure}[htp]
\begin{minipage}[c]{0.9\textwidth}
\tiny{(a)}\includegraphics[width=6.5cm,height=5cm,clip]{fig4a.eps}
\hspace{0.1cm}
\tiny{(b)}\includegraphics[width=6.5cm,height=5cm,clip]{fig4b.eps}
\end{minipage}
\begin{minipage}[c]{0.9\textwidth} 
\tiny{(c)}\includegraphics[width=6.5cm,height=5cm,clip]{fig4c.eps}
\hspace{0.1cm}%
\tiny{(d)}\includegraphics[width=6.5cm,height=5cm,clip]{fig4d.eps}
\end{minipage}
\caption{\label{ipde-x2}(Color online) Plots of (a) $q_u (x,\zeta,b)$ 
vs $x$
for fixed values of $b$ and $\zeta$ and for parameters as in set I,
(b) same as in (a) but for $d$ quark, (c) same as
in (a) but for parameters as in set II, (d) same as in (b) but for 
parameters as in set II.  $b$ is in
GeV$^{-1}$.}

\end{figure}
\begin{figure}[htp]
\begin{minipage}[c]{0.9\textwidth}

\tiny{(a)}\includegraphics[width=6.5cm,height=5cm,clip]{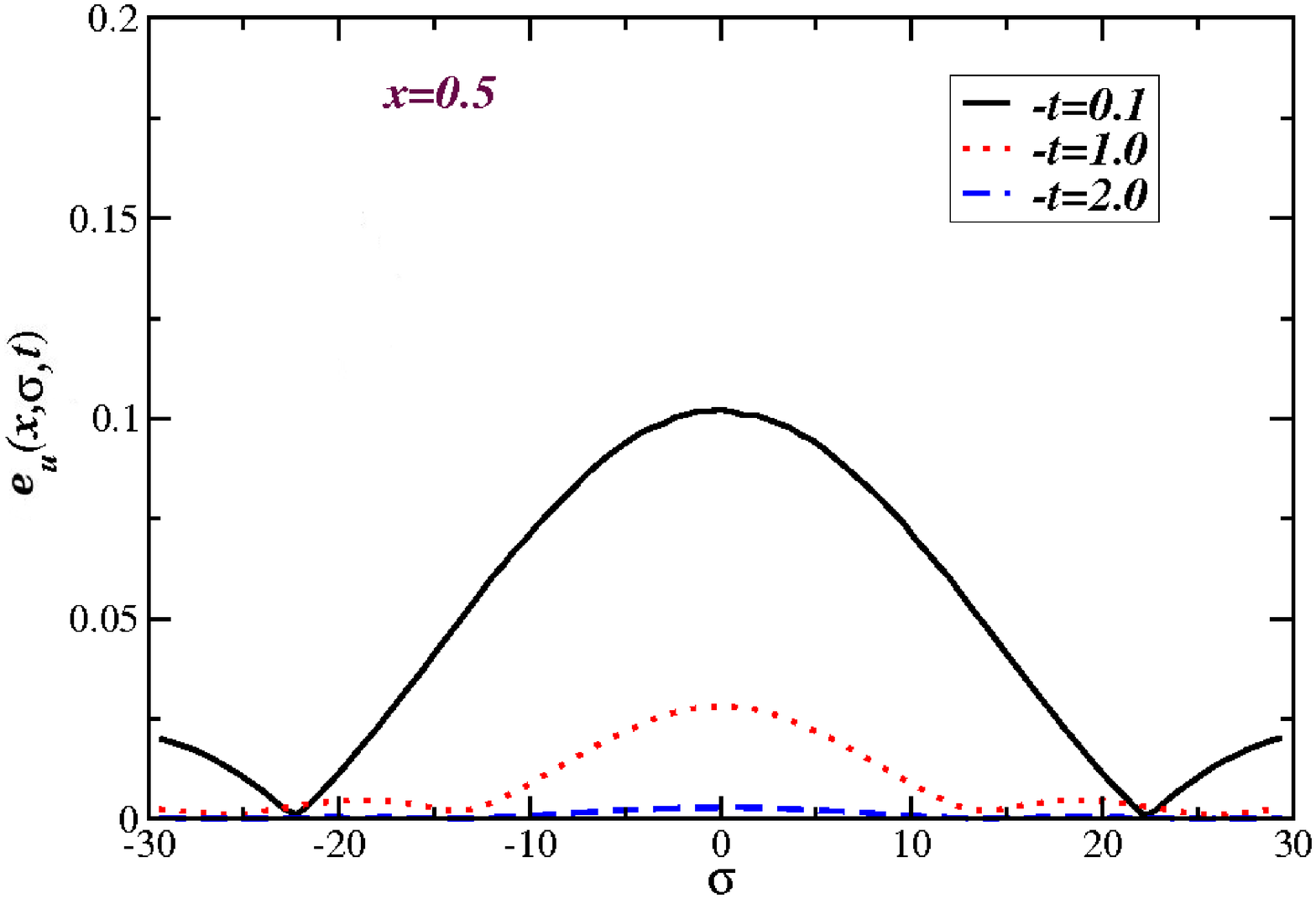}
\hspace{0.1cm}
\tiny{(b)}\includegraphics[width=6.5cm,height=5cm,clip]{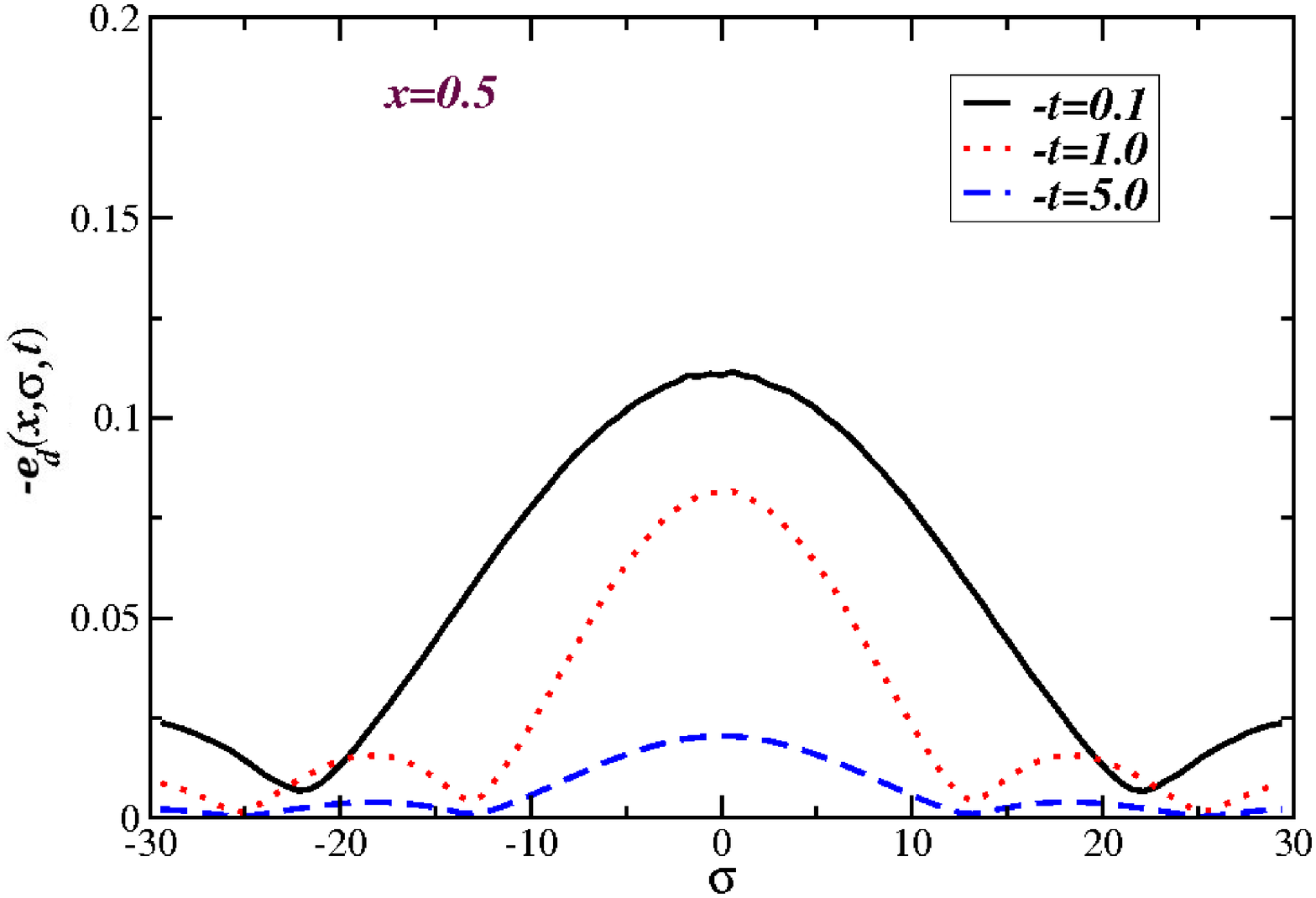}
\end{minipage}
\begin{minipage}[c]{0.9\textwidth} 
\tiny{(c)}\includegraphics[width=6.5cm,height=5cm,clip]{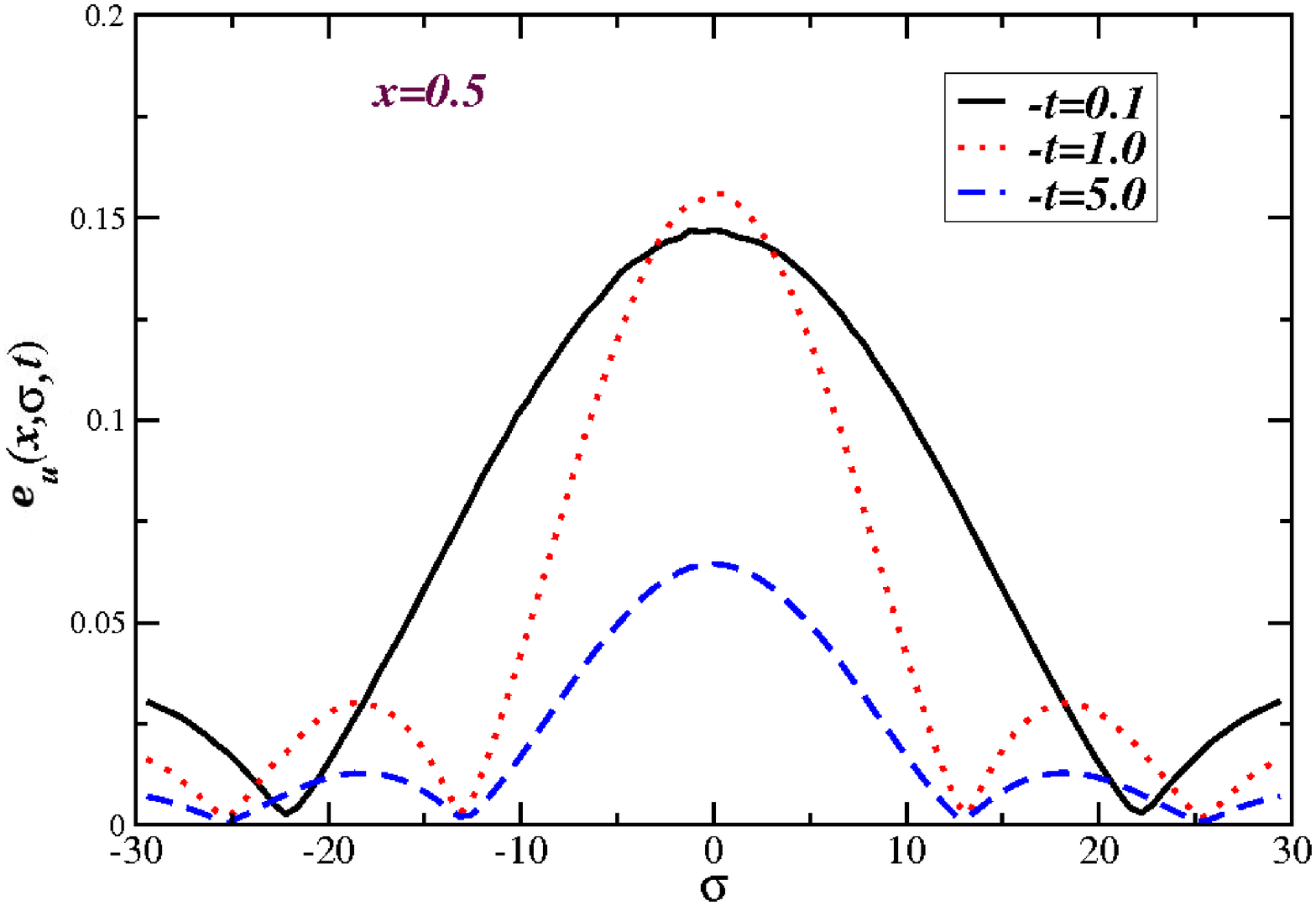}
\hspace{0.1cm}%
\tiny{(d)}\includegraphics[width=6.5cm,height=5cm,clip]{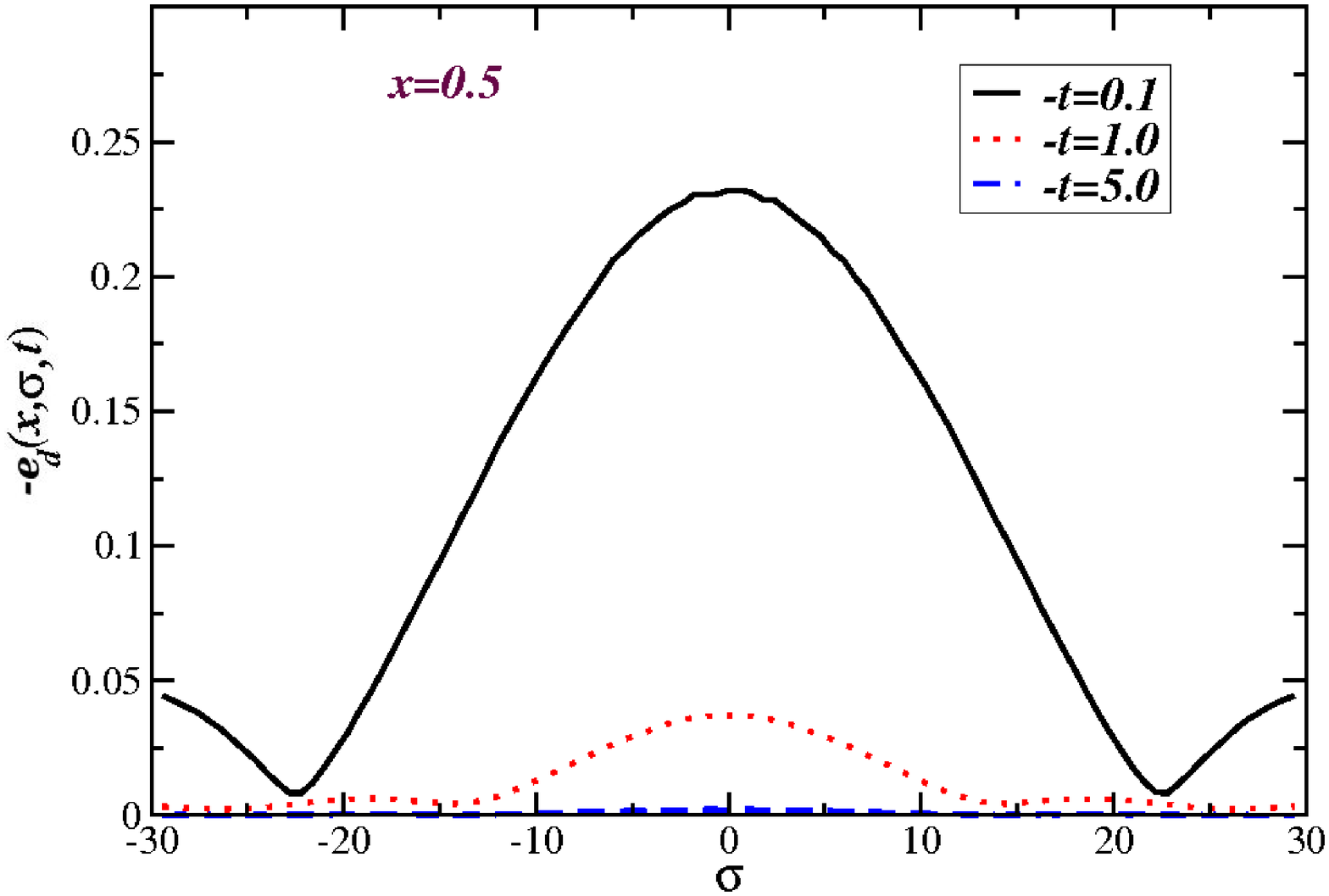}
\end{minipage} 
\caption{\label{E_ud} (Color online) Plots of (a) $e_u (x,\sigma,t)$ vs $\sigma$ , 
(b)  $-e_d (x,\sigma,t)$ vs $\sigma$ for a fixed value of $x$ with different $-t$
in GeV$^2$. (c) same as (a) and (d) same as (b) but for the second parametrization.}
\end{figure}
\begin{figure}[htp]
\begin{minipage}[c]{0.9\textwidth}

\tiny{(a)}\includegraphics[width=6.5cm,height=5cm,clip]{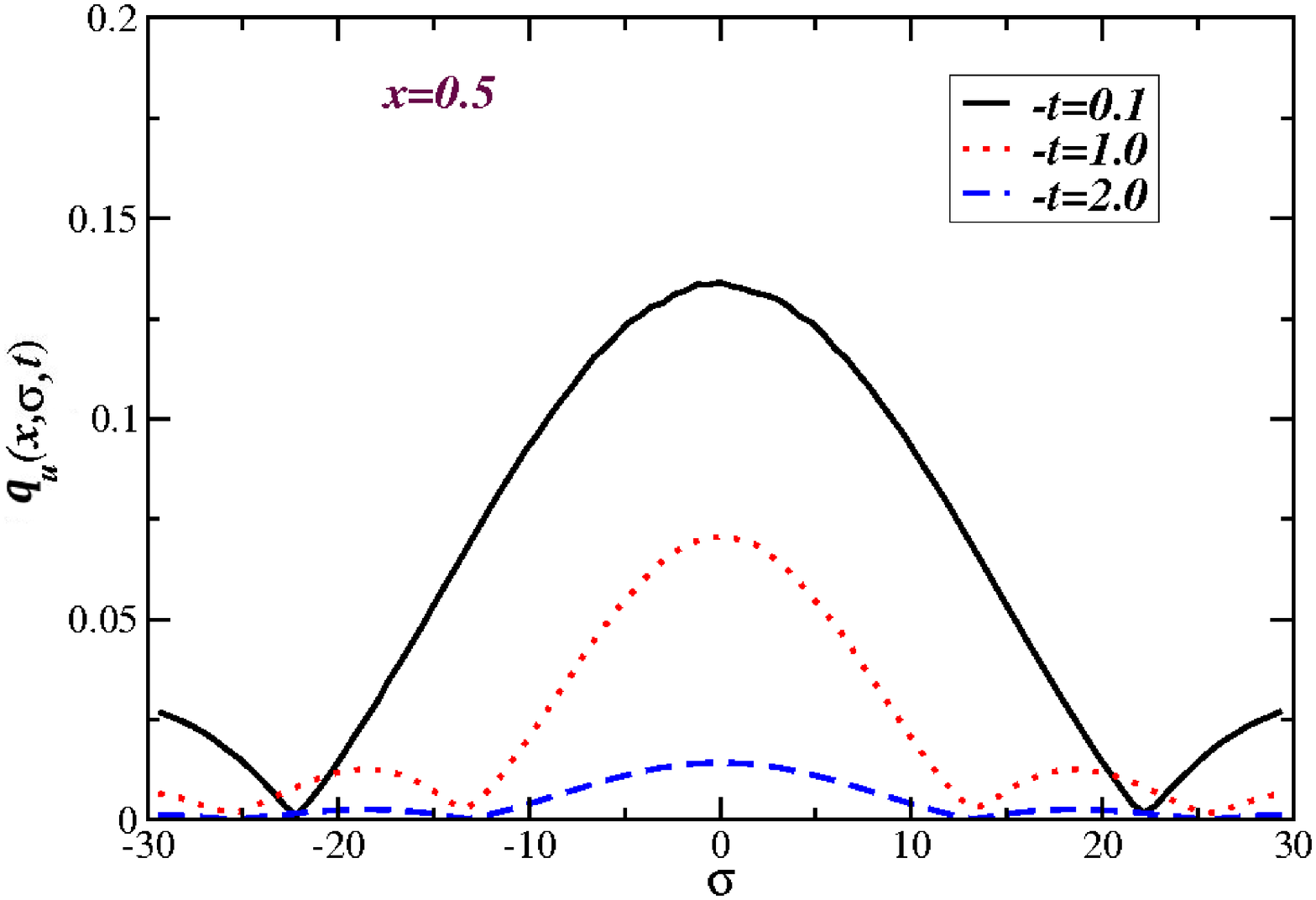}
\hspace{0.1cm}
\tiny{(b)}\includegraphics[width=6.5cm,height=5cm,clip]{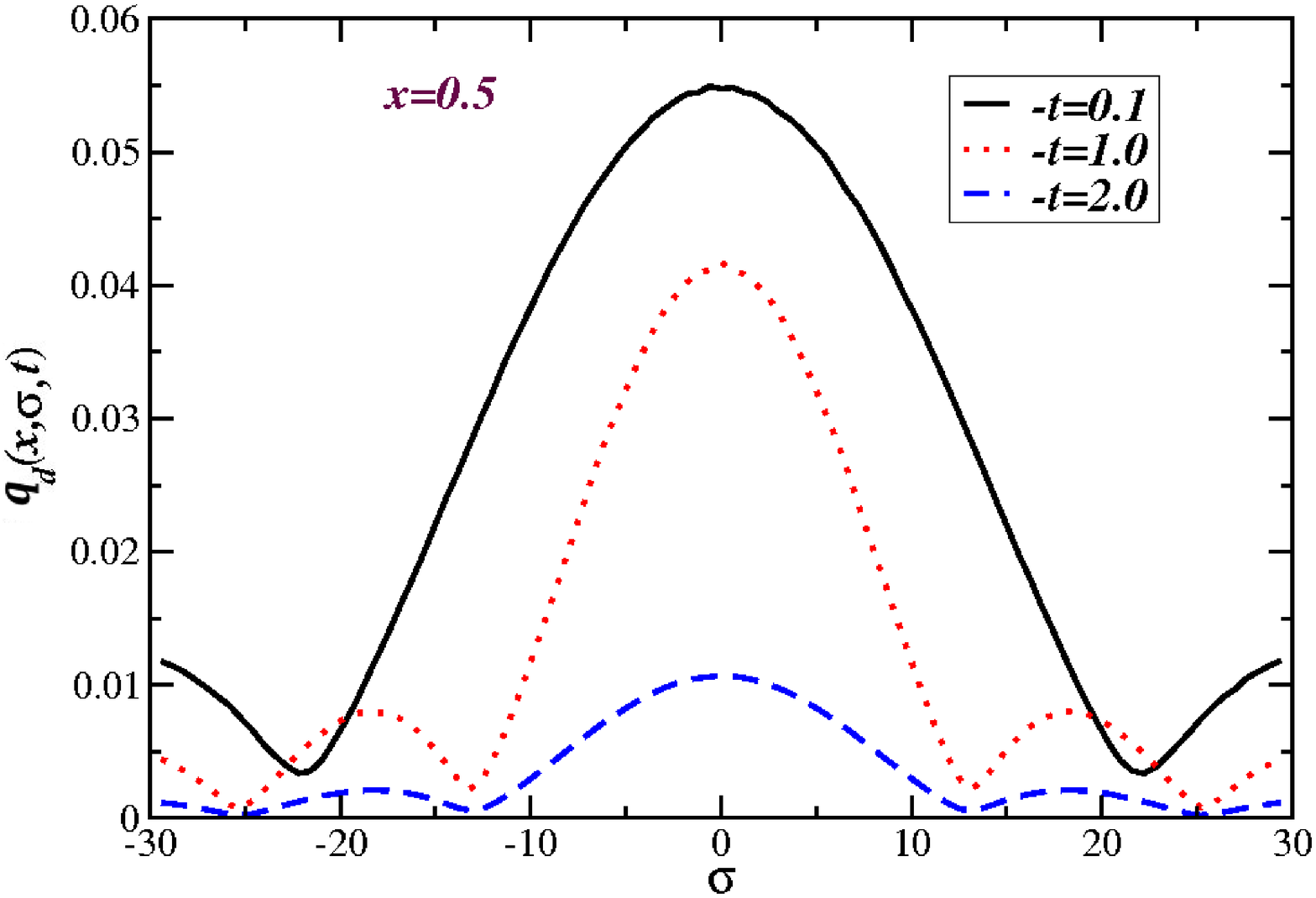}
\end{minipage}
\begin{minipage}[c]{0.9\textwidth} 
\tiny{(c)}\includegraphics[width=6.5cm,height=5cm,clip]{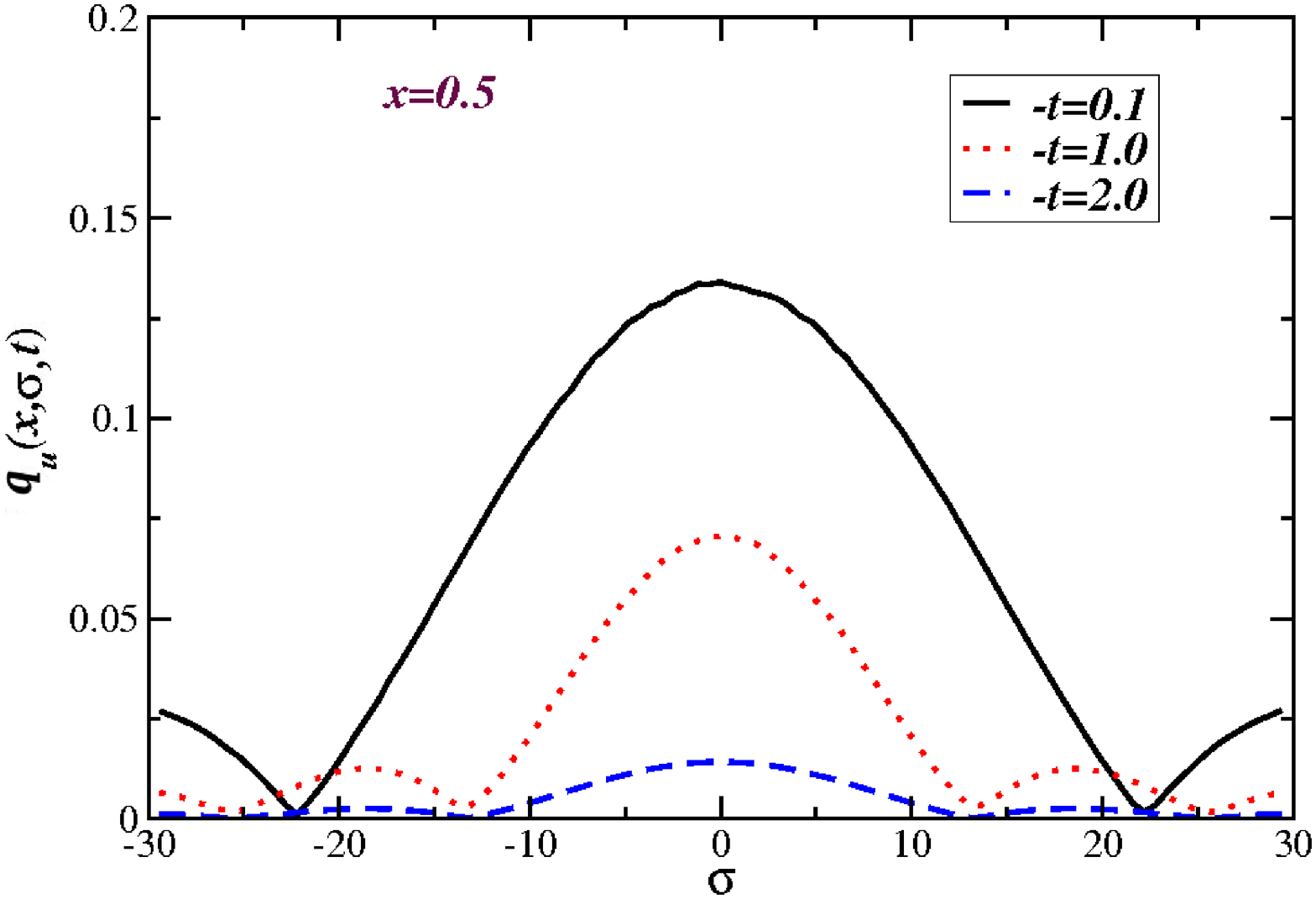}
\hspace{0.1cm}%
\tiny{(d)}\includegraphics[width=6.5cm,height=5cm,clip]{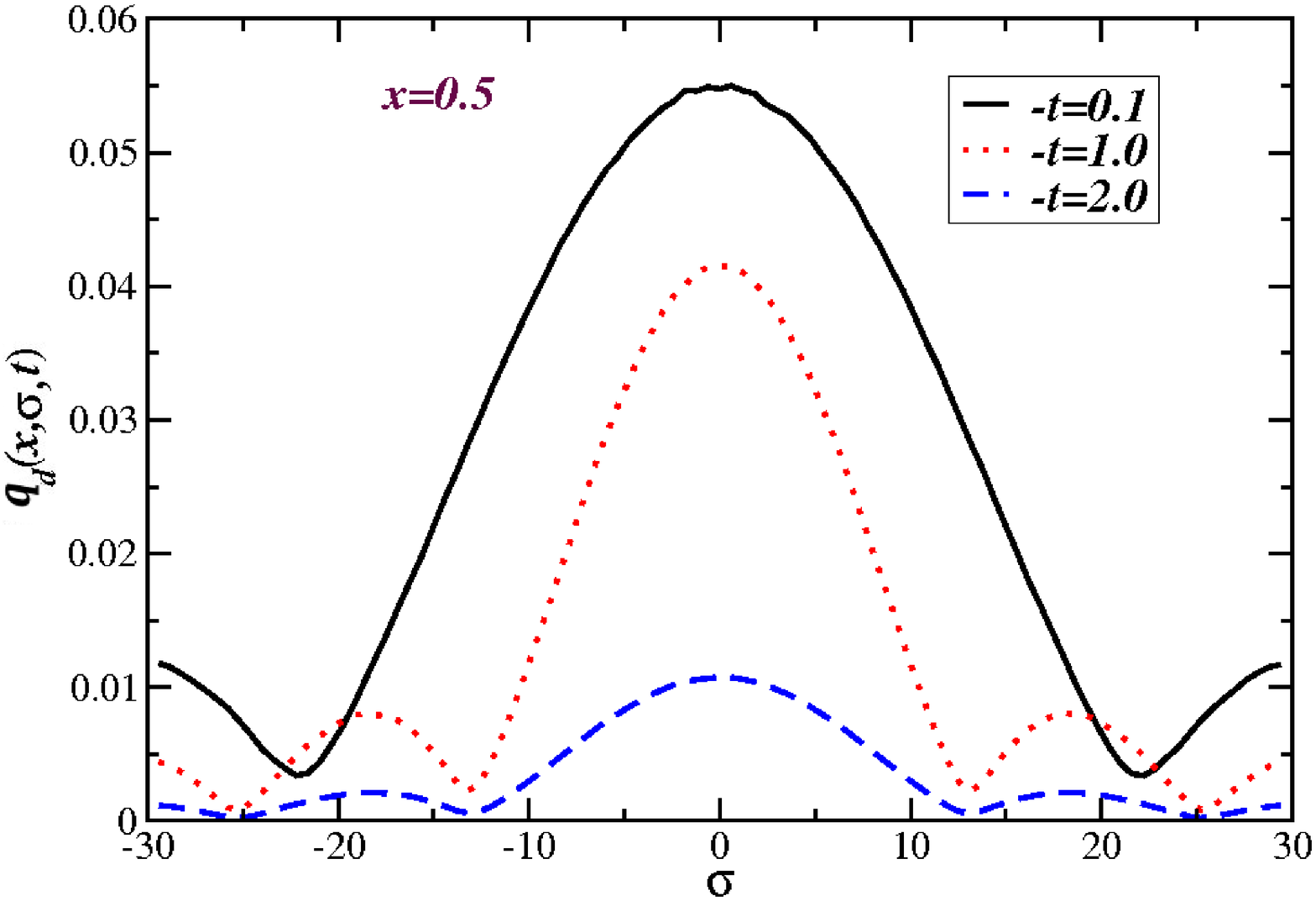}
\end{minipage} 
\caption{\label{H_ud} (Color online) Plots of (a) $q_u (x,\sigma,t)$ vs $\sigma$ , 
(b)  $q_d (x,\sigma,t)$ vs $\sigma$ for fixed values of $x$ 
and different $-t$ in GeV$^2$. (c) same as (a) and (d) same as (b) but for the second 
parametrization.}
\end{figure}
\begin{figure}[htp]
\includegraphics[width=6.5cm,height=5cm,clip]{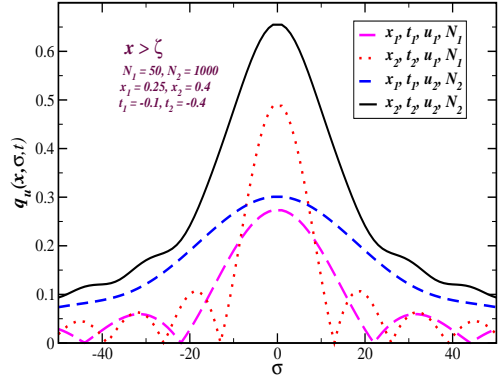}
\caption{\label{model12} (Color online) Plots of $q_u (x,\sigma,t)$ 
in the model given by Eq. (\ref{model1}) (dotted and long dashed) and Eq. (\ref{model2})
(solid and short dashed)}
\end{figure}

\end{document}